\documentclass[twocolumn,           
               showpacs,            
               preprintnumbers,     
               aps,                 
               prd,                 
               a4paper,             
               superscriptaddress,  
               nofootinbib,         
               tightenlines,        
               floats,floatfix      
               ]{revtex4}
\usepackage{graphicx}
\usepackage{latexsym}
\usepackage{amsmath,amssymb}        
\usepackage[draft=false]{hyperref}

\def\be{\begin{equation}}
\def\ee{\end{equation}}
\def\bea{\begin{eqnarray}}
\def\eea{\end{eqnarray}}

\begin{document}

 \typeout{Prints "DRAFT" on each page; does not show in TeXView}
 \title{Directional statistics for WIMP direct detection II: 2-d read-out}
 \author{Ben Morgan}
\affiliation{Department of Physics, University of Warwick, Coventry, CV4 7AL, United Kingdom} 
 \author{Anne M. Green} 
\affiliation{Department of Physics and
 Astronomy, University of Sheffield, Hicks Building, Hounsfield Road,
 Sheffield, S3 7RH, United Kingdom }

\date{\today} 
\pacs{95.35.+d\hfill astro-ph/yymmddd}
\preprint{astro-ph/yymmddd}
\begin{abstract}

The direction dependence of the WIMP direct detection rate provides a
powerful tool for distinguishing a WIMP signal from possible
backgrounds.  We study the the number of events required to
discriminate a WIMP signal from an isotropic background for a detector
with 2-d read-out using non-parametric circular statistics.  We also
examine the number of events needed to i) detect a deviation from
rotational symmetry, due to flattening of the Milky Way halo and ii)
detect a deviation in the mean direction due to a tidal stream. If the
senses of the recoils are measured then of order 20-70 events
(depending on the plane of the 2-d read out and the detector location) 
will be sufficient to reject isotropy of the raw recoil angles at 
90$\%$ confidence.  If the
senses can not be measured these number increase by roughly two orders
of magnitude (compared with an increase of one order of magnitude for
the case of full 3-d read-out). The distributions of the reduced
angles, with the (time dependent) direction of solar motion
subtracted, are far more anisotropic, however, and if the isotropy tests
are applied to these angles then the numbers of events required are
similar to the case of 3-d read-out.
A deviation from rotational symmetry 
will only be
detectable if the Milky Way halo is significantly flattened. The
deviation in the mean direction due to a tidal stream is potentially
detectable, however, depending on the density and direction of the
stream. The meridian plane (which contains the Earth's spin axis) is,
for all detector locations, the optimum read-out plane for rejecting isotropy.
However read-out in 
this plane can not be used for detecting flattening of the 
Milky Way halo or a stream with direction perpendicular to the galactic plane.
In these cases the optimum read-out plane depends on the detector location.

\end{abstract} \maketitle

\section{Introduction}
\label{intro}

Weakly Interacting Massive Particle (WIMP) direct detection
experiments aim to directly detect non-baryonic dark matter via the
elastic scattering of WIMPs on detector nuclei, and are presently
reaching the sensitivity required to detect neutralinos (the lightest
supersymmetric particle and an excellent WIMP candidate). The
direction dependence of the event rate due to the Earth's
motion~\cite{dirndep} provides a powerful WIMP `smoking gun';  a
directional detector needs only of order ten events to differentiate a
WIMP signal from isotropic
backgrounds~\cite{copi:krauss,lehner:dir,pap1}.  Designing a detector
capable of measuring the directions of sub-100 keV nuclear recoils is
a considerable challenge, however. Low pressure gas time projection
chambers (TPCs), such as DRIFT ({\bf D}irectional {\bf R}ecoil {\bf
I}dentification {\bf F}rom {\bf T}racks)~\cite{drift,sean:drift} and
NEWAGE~\cite{newage}, seem to offer the best prospects for a workable
detector.

In paper I~\cite{pap1} we studied the number of events required to
reject isotropy (and hence detect a WIMP signal) and reject rotational
symmetry (and detect flattening of the Milky Way halo) for a range of
observationally motivated halo models, taking into account the detector
response. We also calculated the number of events required to detect a
deviation in the mean direction from the direction of solar motion due
to a tidal stream.  We found that if the senses (i.e the signs) of the
recoil vectors are known then of order ten events will be sufficient to
distinguish a WIMP signal from an isotropic background for all of the
halo models considered, with the uncertainties in reconstructing the
recoil direction only mildly increasing the required number of
events. If the senses of the recoils are not known then the number of
events required is an order of magnitude larger, with a large
variation between halo models, and the recoil resolution is now an
important factor.  The rotational symmetry test required of order a
thousand events to distinguish between spherical and significantly
triaxial halos, however a deviation of the peak recoil direction from
the direction of the solar motion due to a tidal stream could be
detected with of order a hundred events, regardless of whether the
sense of the recoils is known.  While technologies for 3-d TPC
readouts with sufficient resolution to reconstruct sub-100 keV recoils
in 3-d exist~\cite{newage,3dtpc}, the cost and technological challenge of
scaling these up to large, low background WIMP detectors is
considerable.  Therefore in this paper we repeat our analysis for a
detector with less complex 2-d read-out to assess the effects this would have
on the detection potential. Our goals are to assess the 
capabilities of the next generation of detectors and present analysis
techniques which can be applied to real data (taking into account 
experimental practicalities/limitations and the uncertainty in the 
underlying WIMP distribution).

In Sec.~\ref{recoil} we briefly review our calculation of the nuclear
recoil spectrum, including the modelling of the Milky Way halo.  In
Sec.~\ref{stat} we then apply an array of statistical tests aimed at
probing the isotropy (Sec.~\ref{iso}), rotational symmetry
(Sec.~\ref{rotsymtest}) and mean direction (Sec.~\ref{dirtest}) of a
putative WIMP directional signal, before concluding with discussion of
our results in Sec.~\ref{discuss}. In Appendix A we outline the
circular statistics used.

\section{Calculating the nuclear recoil spectrum}
\label{recoil}

The nuclear recoil spectrum depends sensitively on the (unknown) local
WIMP velocity distribution.  Observations and simulations of halos
indicate that it is likely that the Milky Way (MW) halo is triaxial,
anisotropic and contains substructure (see Paper I~\cite{pap1} for
discussion and references) and these properties can lead to
interesting features in the recoil distribution
spectrum~\cite{copi:krauss,radon,pap1}. A generic feature of triaxial
halo models is a flattening of the recoil distribution towards the
Galactic plane, so that the recoil distribution is not symmetric about
the direction of motion of the Sun, $(l_{\odot},b_{{\odot}})$.  WIMPs
from a tidal stream, with velocity dispersion small compared with its
bulk velocity, produce a recoil distribution peaked in the hemisphere
whose pole points in the direction of the stream velocity.  The net
(stream plus smooth background WIMP distribution) peak direction
depends on the direction of the stream and the fraction of the local
density that it contributes.

We consider three fiducial halo models (selected from the twelve
considered in paper I) with properties at the extreme/optimistic end
of the range of expected properties. Model A (1 in paper I) is the
standard halo model, which has a Maxwellian local velocity
distribution with velocity dispersion equal to $270 \, {\rm km \,
s^{-1}}$. Model B (3) is the logarithmic ellipsoidal
model~\cite{evans}, which has a multivariate gaussian velocity
distribution, with shape parameters $p=0.9, q=0.8$ (corresponding to a
density distribution with axis ratios $1:0.78:0.48$) and velocity
anisotropy $\beta=0.4$. Model C (12) is the standard halo model plus a
tidal stream with bulk velocity, in Galactic co-ordinates, $(-65.0, \,
135.0 \, -250.0) \, {\rm km \, s^{-1}}$ and velocity dispersion $30 \,
{\rm km \, s^{-1}}$ comprising $25\%$ of the local
density~\cite{helmi,gondolo:sgr2}.

We calculate the recoil distribution for each halo model via Monte
Carlo simulation, as described in section IIB of Paper
I~\cite{pap1}~\footnote{The recoil momentum spectrum could also be
calculated analytically using the radon transform~\cite{radon},
however for our application Monte Carlo simulations would still be
required to calculate the distributions of the statistics as a
function of number of events.}.  In a realistic directional detector
it will not be possible to measure the nuclear recoil direction with
infinite precision due to multiple scattering and diffusion.  In paper
I~\cite{pap1} we took these effects into account assuming a TPC
detector filled with 0.05 bar CS$_2$, a 10cm drift length over which a
uniform drift field of 1${\rm kV cm}^{-1}$ was applied and a 200
$\mu$m 3-d pixel readout.  Extending these simulations to determine
the angular resolution of a 2-d detector is however complicated;
projection effects combined with multiple scattering of the recoil
will make the resolution a function of the primary recoil direction
and energy. We therefore assume a 0.05bar CS$_2$ TPC detector with a
2-d readout of perfect resolution to provide benchmark figures for the
numbers of events required for the detection of a WIMP signal with a
2-d detector. The numbers of events we find will hence be lower limits
on the numbers required by a real detector. We further take a recoil
threshold of 20 keV, as below this energy the recoil tracks are too
short for their directions to be reconstructed even in our model 3-d
detector.

\begin{figure}[t!]
\includegraphics[width=8.5cm]{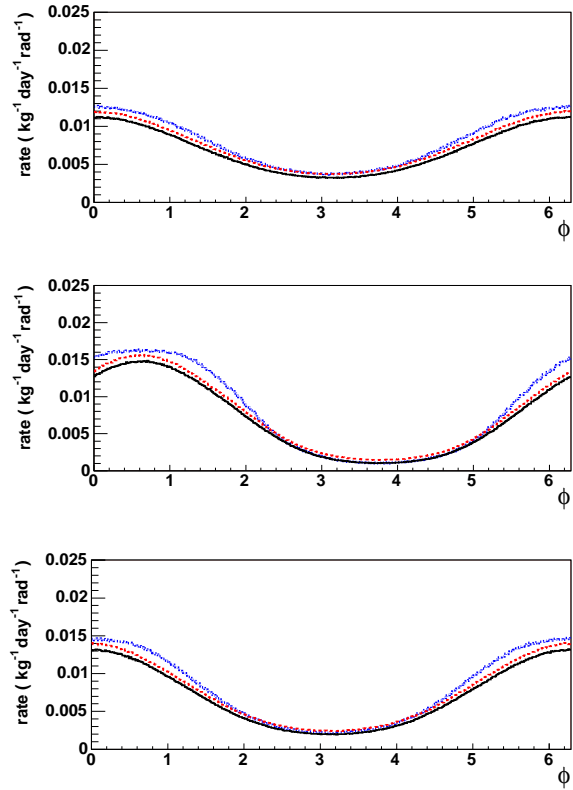}
\caption{The distribution of the raw 2-d angles of the recoils in the horizon, 
meridian and east-west planes from top to bottom 
for a detector located at Boulby
for the three benchmark halo 
models (from bottom to top at $\phi=0$): model A standard halo (black solid 
line), model B triaxial (red dashed),
and model C standard halo plus stream (blue dotted). 
We have set
$\sigma_{0}=10^{-6} \, {\rm pb}$ and $\rho_{0}=0.3 \, {\rm GeV  cm^{-3}}$
and the integrated 
distributions give the total event rate per kilogram, per day 
($0.043, 0.047$ and $0.050 \, {\rm kg}^{-1} {\rm day}^{-1}$ respectively). 
}
\label{rawphidist}
\end{figure}

\begin{figure}[t!]
\includegraphics[width=8.5cm]{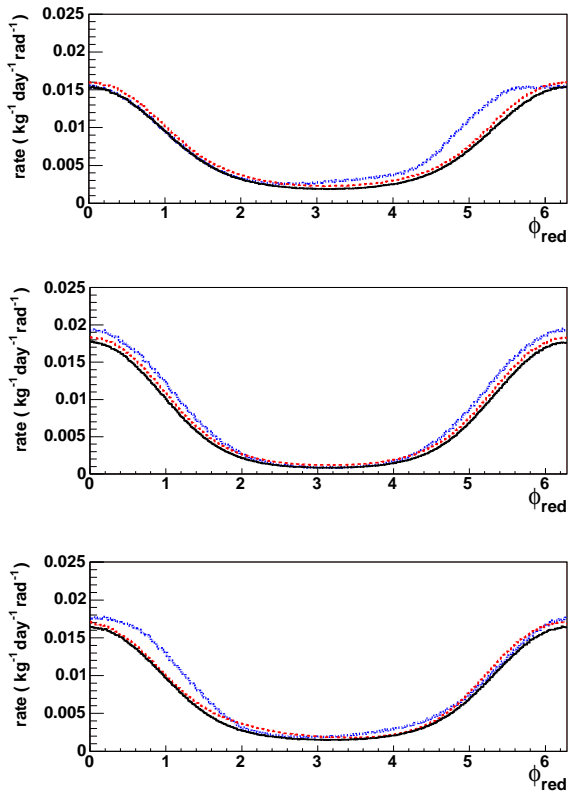}
\caption{As Fig.~\ref{rawphidist} for the reduced angles, 
with the direction of 
motion of the Sun (which is a function of time) subtracted.}
\label{redphidist}
\end{figure}

The orientation of the 2-d readout plane is likely to be
determined by the geometry of the lab.
We therefore define a Cartesian coordinate system
fixed in the laboratory in which the x-axis points towards geographic
north, the y-axis towards geographic west and the z-axis towards the
zenith. The three simplest possible 2-d
read-out planes in this frame are: 
meridian (x-z plane), horizon (x-y plane) and
east-west (y-z plane). We measure 2-d directions by projecting the 3-d
recoil vectors into each of these planes and measuring the
(right-handed) azimuthal angle $\phi$ between the projected vector and
the z-axis in the meridian plane, the x-axis in the horizon plane and
the z-axis in the east-west plane. The orientation of the read-out planes
also depends on the latitude of the detector.
We focus mainly on a detector at the Boulby
mine, where the DRIFT detector is currently located,
which is at a latitude of $54.5^{\circ}$ N.  Potential locations
are limited by the requirement of a suitable deep underground
laboratory (in order to shield cosmic-ray backgrounds). All of the
proposed directional detector locations which we are aware of lie at
mid-northern latitudes (e.g. Kamioka, SnoLab). To cover the range of
possible locations we also examine detectors located at $36.5^{\circ}$ N
(Kamioka) and $46.1^{\circ}$ N (Sudbury).

The distributions of the $\phi$ angles in each plane for each halo
model are generated from the 3-d recoil distributions in the Galactic
frame by Monte Carlo simulation of the sidereal time dependent
coordinate transform to the detector frame, together with the 2-d
projection procedure described above. The size of the anisotropy will
be largest if the maximum in the recoil direction distribution
a) spends as much time as
possible close to the read-out plane (this minimises the spread in the
2-d distribution caused by projection effects) and b) has minimal
motion in $\phi$ (this minimises the spread due to
time-averaging). For smooth WIMP distributions the peak recoil
direction is the direction of Solar motion and these requirements are
fulfilled for any plane whose normal is at $90^{\circ}$ to the spin
axis of the Earth, or equivalently which contains the Earth's spin
axis (see also Ref.~\cite{copi2d}).  This is the case for all meridian
planes and so the anisotropy should be largest in this plane for any location.

In Fig.~\ref{rawphidist} we plot the raw 2-d angle distributions for
the three benchmark halo models for each read-out plane for a detector
located at Boulby. For normalisation purposes we have taken the
WIMP-nucleon cross section and the local WIMP density to be
$\sigma_{0}=10^{-6} \, {\rm pb}$ and $\rho_{0}=0.3 \, {\rm GeV
cm^{-3}}$ respectively. As expected the peak-to-trough variation is
largest in the meridian plane. The peak-to-trough variation is smallest
in the horizon plane as, at Boulby, this plane is furthest from the
Earth's spin axis. The standard halo and standard halo with stream
(models A and C) have similar peak-to-trough variations, while the
triaxial model has a smaller variation (this is not so obvious from
the plot, as the three models have different normalisations,
reflecting the different event rates above the $20 \, {\rm keV}$
threshold).  This suggests that it will be hardest to reject isotropy
for read out in the horizon plane and/or the triaxial halo model.  
The angle distributions at Sudbury and Kamioka are qualitatively similar
with the peak-to-trough variation remaining constant in the meridian
plane and increasing (decreasing) in the horizon (east-west) plane as
the detector location is moved South.

A major difference from the 3-d analysis is that here the recoil
directions cannot be transformed from the lab rest frame to the
Galactic rest frame. 
In the 3-d case, the time dependent conversion
between the lab and Galactic coordinate frames tends to wash out any
anisotropies in lab backgrounds so that they are isotropic in the
Galactic rest frame. The modulation of the mean
recoil direction with sidereal time (e.g. Ref.~\cite{morgannoon})
potentially provides a means of checking the Galactic nature of an
anisotropic 2-d signal observed in the lab frame. However, determining
the mean direction as a function of time necessarily requires large
quantities of data. Instead, we use the direction of solar motion
projected into each plane, $\mu_{\odot}(t)$, which is sidereal 
time-dependent in the lab frame, to calculate the reduced angle, $\phi_{\rm
red}$, of each recoil:
\begin{equation}
\phi_{\rm red} = \phi - \mu_\odot(t) \,,
\end{equation}
where $t$ is the sidereal time at which the recoil occurred.  The
time-dependent nature of the reduced angle transform means that any
anisotropy in lab backgrounds will be washed out to give isotropic
reduced angle distributions. 

The reduced angle distributions for the three benchmark halo models
for a detector located at Boulby are shown in Fig.~\ref{redphidist},
and it can be seen that these peak at $\sim 0^{\circ}$ and have a
higher degree of anisotropy compared with the raw distributions in
Fig.~\ref{rawphidist}. The reduced angle distributions for model C
(with the tidal stream) shows significant excesses with respect to the
distributions for the two smooth halo models (at $\phi_{\rm red} \sim
180^{\circ}-300^{\circ}$, $\sim 0^{\circ} \pm 1^{\circ}$ and $\sim
0^{\circ}-100^{\circ}$ in the horizon, meridian and east-west planes
respectively).  The mean reduced angles are $\bar{\phi}_{\rm
red}=348^\circ$, $<0.1^{\circ}$ and $4.2^{\circ}$ respectively.  The
deviation of the mean reduced angle from zero remains zero in the
meridian plane and increases (decreases) in the horizon (east-west)
plane as the detector is moved South.  The reduced angle distribution
for halo model B has a small asymmetry in the horizon plane at Boulby
(the rate at $\phi$ is smaller (larger) than that at $2 \pi - \phi$
for $\phi < (>) \pi/2$). There is a larger deviation from symmetry in
the horizon and east-west planes at Sudbury and in the horizon plane
at Kamioka. The reduced angle distributions in the meridian plane are
always symmetric, as this plane is perpendicular to the flattening of
the halo.

\begin{figure}[t!]
\includegraphics[width=8.5cm]{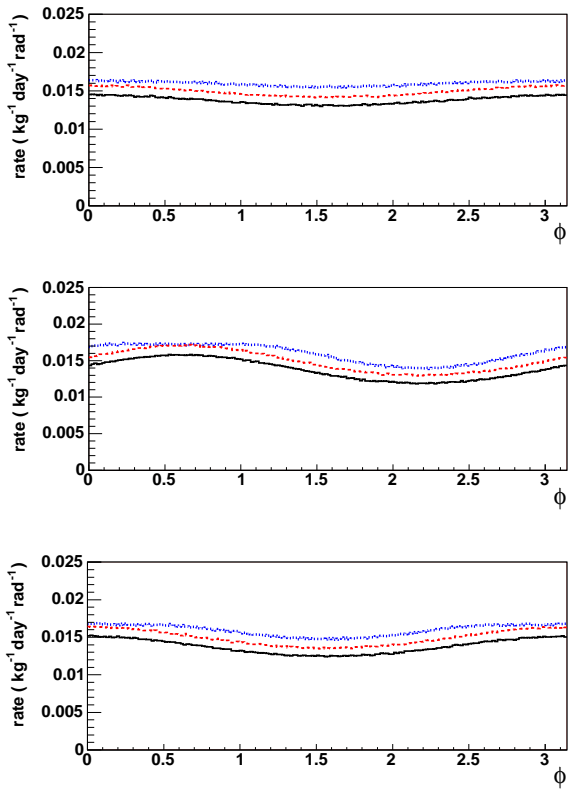}
\caption{The raw angle distributions for axial data, where the sense of the
recoil can not be measured (halo models, read-out planes and detector location
as in Fig.~\ref{rawphidist}).}
\label{rawphidistaxial}
\end{figure}

\begin{figure}[t!]
\includegraphics[width=8.5cm]{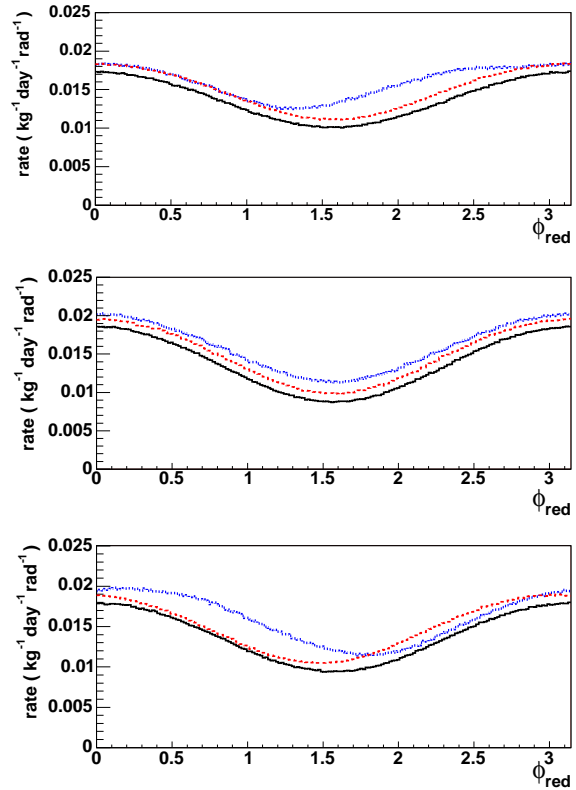}
\caption{The reduced angle distributions for axial data, where the sense of the
recoil can not be measured (halo models, read-out planes and detector location as in Fig.~\ref{rawphidist}).}
\label{redphidistaxial}
\end{figure}

It is possible that the absolute signs of the recoil vectors
(i.e. their `sense' $+\vec{x}$ or $-\vec{x}$) may not be measurable,
The 2-d raw and reduced angle distribution at Boulby in this case are
plotted in Figs.~\ref{rawphidistaxial} and \ref{redphidistaxial}.  The
peak-to-trough variations of both the raw and reduced angles are
significantly smaller than the corresponding vector angle
distributions. The anisotropy of the raw axial angle distributions are
very small and in particular the direction of the stream in model C is
such as to produce an almost flat raw angle distribution in the
horizon plane. This is also the case at Kamioka in the east-west plane.

\section{Applying Statistical tests}
\label{stat} 

As we discussed more extensively in Paper I~\cite{pap1} a WIMP search
strategy with a directional detector can be divided into three regimes
i) search (aiming to detect a non-zero signal), ii) confirmation (does
the signal have the anisotropy expected for a Galactic signal?) iii)
exploitation (extraction of information about the local WIMP velocity
distribution, for instance flattening of the halo or the presence 
of a tidal stream). We therefore consider three simple hypotheses to test
\begin{enumerate}
\item Is the recoil direction distribution uniform?
\item Is the recoil distribution rotationally symmetric about the direction
of solar motion?
\item Does the mean direction deviate from the direction of solar motion?
\end{enumerate}
aimed at detecting a WIMP signal, flattening of the MW halo and the presence
of a stream of WIMPs respectively.

The recoil directions projected into the read-out plane constitute
2-d vectors, or, if the senses are  not known, undirected lines or axes, and
so can equivalently be represented as points on a circle, parameterised
by their angle relative to some fixed point/direction. This allows
us to use statistical inference methods developed for the analysis of
circularly distributed data (for a review of this extensive field see
the standard texts such as Refs.~\cite{mardia:jupp,fisher}).  We investigate
a variety of non-parametric statistics designed to test the isotropy
(Secs.~\ref{iso}), rotational symmetry (Sec.~\ref{rotsymtest}) and
mean direction (Sec.~\ref{dirtest}) for the benchmark halo
models discussed in Sec.~\ref{recoil}. 

For each halo model and readout plane we calculate the 
probability distribution of each test statistic, for a given number of 
events $n$, by Monte Carlo generating $10^5$ experiments each observing $n$ 
recoils drawn from our calculated $\phi$ distributions.
We then compare this with the null distribution of the
statistic, under the assumption of isotropy/rotational symmetry/zero
mean reduced angle calculated using analytic expressions where available
and otherwise via Monte Carlo simulation. Specifically we calculate
the rejection and acceptance factors, $R$ and $A$, for each value of
the statistic. The rejection factor is the probability of measuring a
smaller absolute value of the statistic if the null hypothesis is
true or equivalently the confidence with which the null hypothesis can
be rejected given that measured value of the statistic.  The
acceptance is the probability of measuring a larger absolute value of
the test statistic if the alternative hypothesis is true or
equivalently the fraction of experiments in which the alternative
hypothesis is true that measure a larger absolute value of the test
statistic and hence reject the null hypothesis at confidence level
$R$.  Clearly a high rejection factor is required to reject the null
hypothesis.  We also require a high acceptance, otherwise any one
experiment might not be able to reject the null hypothesis at the
given rejection factor or the null hypothesis might be erroneously
rejected. We therefore find, using a search by bi-section, the number
of events required for $A_{\rm c}=R_{\rm c}=0.9$ and $0.95$. For
further details on this procedure see Appendix C of Paper
I~\cite{pap1}.

With 2-d axial data the conventional (but not rigorously justified)
procedure is to double the angles of the 2-d vectors and then
reduce them modulo $360^{\circ}$ (see Sec.~\ref{intro:A} and
Ref.~\cite{fisher}) before applying the statistical tests for circular
data. This procedure can be used in the case of the isotropy tests
but not the symmetry and mean direction tests, as the effect of this
transformation on the spread of the reduced angles is not straight
forward.

\begin{table}[htbp]
\begin{tabular}{|c|c|c|c||c|c|c|}
\hline
Halo & \multicolumn{3}{c||}{ 
 $( R_{\rm c}, A_{\rm c})=(0.90,0.90)$} &
\multicolumn{3}{c|}{ 
 $( R_{\rm c}, A_{\rm c})=(0.95,0.95)$}
\\
\cline{2-7}
Model &  Rayleigh & Kuiper & Watson 
& Rayleigh & Kuiper & Watson
\\
\hline \hline
 & \multicolumn{6}{|c|} {Horizon plane} \\
\hline
A & 62 & 70 & 63 & 92 & 103 & 92\\
B & 70 & 78 & 71& 103 & 115 & 104 \\
C &  61 & 70 & 62 & 91 & 101 & 91 \\
\hline
 & \multicolumn{6}{c|} {Meridian plane} \\
\hline
A & 18 & 20 & 17 & 26 & 29 & 25 \\
B & 21 & 23 & 20   & 30 & 33 & 29    \\
C & 16 & 18 & 16  & 24 & 26 & 23     \\
\hline 
 & \multicolumn{6}{c|} {East-west plane} \\
\hline

A & 28 & 31 & 28 & 41 & 46 & 41\\
B & 32 & 36 & 32  & 47 & 53 & 47   \\
C & 28 & 31 & 28  & 41 & 46 &  41   \\
\hline

\end{tabular}
\caption{Number of recoil events
required to reject isotropy of recoil directions, 
$N_{\rm iso}$, at $90  \, (95)\%$
confidence in $90 \, (95)\%$ of experiments 
for the Rayleigh, Kuiper and Watson statistics (Appendix A2)
for fiducial halo models A (standard), B (triaxial) and 
C (standard plus stream) for each read-out plane for a 
detector located at Boulby.}
\label{neventsiso}
\end{table}

Throughout we assume zero background. This is a reasonable expectation
for the next generation of experiments made from low activity
materials with efficient gamma rejection and
shielding, located deep underground
~\cite{drift2:design}.  As with the statistical tests of
paper I, non-zero backgrounds can be incorporated into the 2-d tests
as described in Section III of paper I. This  requires a known
background rate, so given that the next generation of directional
detectors expect to have essentially zero background, we assume zero
background to provide benchmark figures.  We should also emphasise
that these non-parametric tests, unlike likelihood analyses, do not make any
assumptions about the form of the recoil spectrum and can hence be 
applied to real data (where the underlying local WIMP velocity 
distribution is unknown).

\subsection{Tests of isotropy}
\label{iso}

We initially focus on  a detector located at Boulby.
For the Rayleigh, Kuiper and Watson statistics described in
Appendix~\ref{iso:A} we determine the minimum number of events
required to reject isotropy of recoil directions at $90 \, (95) \%$
confidence in $90 \, (95)\%$ of experiments (i.e. for rejection and
acceptance probabilities of $ R_{\rm c} =A_{\rm c}=0.9$ and $0.95$),
$N_{\rm iso}$, for all three benchmark halo models. The results for the
raw angle distributions are
tabulated in Tables~\ref{neventsiso} and \ref{neventsiso2} for vector
and axial data respectively.

\begin{table}[htbp]
\begin{tabular}{|c|c|c|c||c|c|c|}
\hline
Halo & \multicolumn{3}{c||}{ 
 $( R_{\rm c}, A_{\rm c})=(0.90,0.90)$} &
\multicolumn{3}{c|}{ 
 $( R_{\rm c}, A_{\rm c})=(0.95,0.95)$}
\\
\cline{2-7}
Model &  Rayleigh & Kuiper & Watson 
& Rayleigh & Kuiper & Watson
\\
\hline \hline
 & \multicolumn{6}{|c|} {Horizon plane} \\
\hline
A & 8 300  & 9 300  & 8 400  & 12 000  & 14 000  & 12 000 \\
B & 10 000 & 11 000  & 10 000 & 15 000   & 16 000   &  15 000  \\
C & $>20 000$   &  $>20 000$  &  $>20 000$  &  $>20 000$  &  $>20 000$  & 
 $>20 000$  \\
\hline
 & \multicolumn{6}{c|} {Meridian plane} \\
\hline
A & 1 100 & 1 200  & 1 100  & 1 600  & 1 800  & 1 600  \\
B & 1 200 & 1 300  &  1 200   &  1 800 & 2 000 &  1 800    \\
C & 1 700 & 1 900 &  1 800  & 2 600 & 2 800 & 2 600     \\
\hline 
 & \multicolumn{6}{c|} {East-west plane} \\
\hline
A & 2 100 & 2 800 & 2 200  & 3 100  & 3 500  & 3 200 \\
B & 2 400 &  2 800 &  2 400  & 3 600  & 4 000    & 3 600 \\
C & 5 500 & 5 900   &  5 400 &  8 200 & 8 700  &  8 000   \\
\hline
\end{tabular}
\caption{Same as Table I for axial data (numbers quoted to 2 significant 
figures).}
\label{neventsiso2}
\end{table}

\begin{table}[htbp]
\begin{tabular}{|c|c|c|c||c|c|c|}
\hline
Halo & \multicolumn{3}{c||}{ 
 $( R_{\rm c}, A_{\rm c})=(0.90,0.90)$} &
\multicolumn{3}{c|}{ 
 $( R_{\rm c}, A_{\rm c})=(0.95,0.95)$}
\\
\cline{2-7}
Model &  Rayleigh & Kuiper & Watson 
& Rayleigh & Kuiper & Watson
\\
\hline \hline
 & \multicolumn{6}{|c|} {Horizon plane} \\
\hline
A & 21 & 33 & 21 & 30 & 33 & 31\\
B & 24 & 26 & 24& 34 & 38 & 25 \\
C &  26 & 29 & 27 & 38 & 42 & 39 \\
\hline
 & \multicolumn{6}{c|} {Meridian plane} \\
\hline
A & 12 & 14 & 13 & 18 & 20 & 18 \\
B & 14 & 16 & 14   & 20 & 23 & 21    \\
C & 13 & 14 & 13  & 18 & 20 & 19     \\
\hline 
 & \multicolumn{6}{c|} {East-west plane} \\
\hline

A & 17 & 18 & 17 & 24 & 27 & 25\\
B & 19 & 21 & 19  & 28 & 31 & 28   \\
C & 19 & 21 & 19  & 28 & 31 &  28   \\
\hline

\end{tabular}
\caption{Number of events
required to reject isotropy of reduced angles.}
\label{neventsisored}
\end{table}

\begin{table}[htbp]
\begin{tabular}{|c|c|c|c||c|c|c|}
\hline
Halo & \multicolumn{3}{c||}{ 
 $( R_{\rm c}, A_{\rm c})=(0.90,0.90)$} &
\multicolumn{3}{c|}{ 
 $( R_{\rm c}, A_{\rm c})=(0.95,0.95)$}
\\
\cline{2-7}
Model &  Rayleigh & Kuiper & Watson 
& Rayleigh & Kuiper & Watson
\\
\hline \hline
 & \multicolumn{6}{|c|} {Horizon plane} \\
\hline
A & 310  & 320  & 310  & 450  & 460  & 460 \\
B & 360 & 400  & 360 & 520   & 590   &  530  \\
C &  670  &  740  & 670  &  990  & 1100   & 990  \\
\hline
 & \multicolumn{6}{c|} {Meridian plane} \\
\hline
A & 160 & 180  & 170  &  240 & 270  & 240  \\
B & 200 & 220  & 200  & 290   & 320  &  290    \\
C & 280 & 310 & 280   & 410 & 460 & 410      \\
\hline 
 & \multicolumn{6}{c|} {East-west plane} \\
\hline
A & 220 & 240 &  220  & 320
  & 360  & 330  \\
B & 260 & 300  & 260   & 380  & 420   & 380  \\
C & 310 & 350   & 320  & 460  & 510  & 470    \\
\hline
\end{tabular}
\caption{Same as Table III for axial data (numbers quoted to 2 significant 
figures).}
\label{neventsisored2}
\end{table}

For vector data, in the horizon plane roughly 60 (90) events are
required to reject isotropy at $90 \,(95)\%$ confidence in $90 \,
(95)\%$ of experiments. The numbers of events required in the
east-west and meridian planes are smaller by factors of roughly 2 and
3 respectively. For each read-out plane the Rayleigh and Watson tests
are equally powerful, with the Kuiper test requiring roughly $10\%$
more events.  The triaxial model needs $\sim 10\%$ more events than
the standard halo model while the tidal stream in model C does not
effect the number of events required. These quantitative trends match
the expectations from examining the angle distributions in Sec. II.
We note that these number are a factor of $2-7$ (depending on
the read-out plane) larger than for a 3-d detector with perfect
resolution~\cite{pap1}.

The same trends (between halo models, read-out planes and statistics)
are seen for the case of axial data, however the number of events
required are increased by roughly 2 orders of magnitude.  This is
significantly worse than the case of 3-d data, where we found an
increase of a single order of magnitude~\cite{pap1}.  The increase is
even larger for model C in the horizon plane (as noted in Sec. II the
direction of the stream in this case is such as to produce an almost
flat axial angle distribution).

We also apply the isotropy tests to the reduced angle
distributions. The resulting numbers of events required to reject
isotropy are given in Tables~\ref{neventsisored} and
~\ref{neventsisored2} for vector and axial data respectively. The
variations between halo models and statistics are broadly the same as
when the tests are applied to the raw angle distributions, however the
numbers of events required are significantly smaller: $\sim 30\%$ in
the meridian and east-west planes and a factor of $\sim$ 3 in the
horizon plane. The smaller numbers of events reflect the greater
anisotropy of the reduced angle distributions.

\begin{table}[htbp]
\begin{tabular}{|c|c|c|c||c|c|c|}
\hline
Halo & \multicolumn{3}{c||}{ 
 $( R_{\rm c}, A_{\rm c})=(0.90,0.90)$} &
\multicolumn{3}{c|}{ 
 $( R_{\rm c}, A_{\rm c})=(0.95,0.95)$}
\\
\cline{2-7}
Model &  Boulby & Sudbury & Kamioka 
& Boulby & Sudbury & Kamioka
\\
\hline \hline
 & \multicolumn{6}{|c|} {Horizon plane} \\
\hline
A &  21 & 20  & 17 & 30  & 28 & 25 \\
B & 24  & 22 & 19 & 34 & 32 & 28 \\
C &  26 & 24 & 20 & 38 & 35  & 28 \\
\hline
 & \multicolumn{6}{c|} {Meridian plane} \\
\hline
A & 12 & 12 & 12  & 18 & 18 &  18 \\
B & 14 & 14 & 14   & 20 & 21  &  20   \\
C & 13 & 13 & 13  & 18 & 18 &   18   \\
\hline 
 & \multicolumn{6}{c|} {East-west plane} \\
\hline

A & 17  & 19 & 21  & 24 & 28  &  30 \\
B & 19 & 21 &  24 & 28 & 31 &  34  \\
C & 19  & 23 & 26  & 28 & 33 &  38    \\
\hline

\end{tabular}
\label{locationtable}
\caption{Number of recoil events
required to reject isotropy of reduced angles
using the Rayleigh statistic for each halo model
for detectors located at Boulby ($54.5^{\circ}$ N), Sudbury ($46.1^{\circ}$ N) 
and Kamioka ($36.5^{\circ}$ N).}
\end{table}

\begin{table}[htbp]
\begin{tabular}{|c|c|c|c||c|c|c|}
\hline
Halo & \multicolumn{3}{c||}{ 
 $( R_{\rm c}, A_{\rm c})=(0.90,0.90)$} &
\multicolumn{3}{c|}{ 
 $( R_{\rm c}, A_{\rm c})=(0.95,0.95)$}
\\
\cline{2-7}
Model &  Boulby & Sudbury & Kamioka 
& Boulby & Sudbury & Kamioka
\\
\hline \hline
 & \multicolumn{6}{|c|} {Horizon plane} \\
\hline
A & 310  & 260  & 220  & 450  & 390 & 330  \\
B & 360  & 310 & 260  & 520  & 460 & 390 \\
C & 670  & 340 & 320 & 990 & 490  & 470 \\
\hline
 & \multicolumn{6}{c|} {Meridian plane} \\
\hline
A & 160 & 160 & 160  & 240 & 240 & 240  \\
B & 200 & 200 & 200   & 290 & 290  & 290     \\
C & 280 & 280 & 280  & 410 & 410 &   410   \\
\hline 
 & \multicolumn{6}{c|} {East-west plane} \\
\hline

A & 220  & 250 & 300  & 320 & 370  &  440 \\
B & 260 & 300 & 350  & 380 & 440 &  520  \\
C & 310  & 330 & 610  & 460 & 490  & 900     \\
\hline

\end{tabular}
\caption{As table V for axial data.}
\end{table}

Tables V and VI compare the numbers of events required to reject
isotropy of reduced angles using the Rayleigh statistic for detectors
located at Boulby, Sudbury and Kamioka for vectorial and axial data
respectively. As expected from the angle distributions, the number of
events is smallest (and constant) in the meridian plane and decreases
(increases) on moving South for read-out in the horizon (east-west)
plane. The, fractional, variation on moving from Boulby to Kamioka is
larger for the axial angles, $\sim 35\%$ compared with $\sim 20\%$ for
the vectorial angles. The same qualitative trends generally occur for
the raw angles but the variations are larger: a
factor of $\sim 2$ ($\sim 3.4$) for vectorial (axial) data. The one
exception is model C for axial data. At Sudbury isotropy can be
rejected with $<20 \,000$ events in all read-out planes, whereas at
Kamioka (Boulby) more than $>20 \,000$ events are required for the
east-west (horizon) plane reflecting the close to flat raw axial angle 
distributions for these location/read-out plane combinations.

Finally, we translate the numbers of events required to reject isotropy
with a detector located at Boulby 
into the equivalent detector exposures, $E$, required to observe this
number of events. If the senses of the recoil directions are observed,
isotropy of the reduced angle distribution could be rejected at $95\%$
confidence in $95\%$ of experiments for WIMP-nucleon cross-sections
down to $\sigma_0 \sim 7, 4$ and $6 \times10^{-9} \,{\rm pb}$ for
read-out in the horizon, meridian and east-west planes respectively,
with an exposure of $E \sim 10^5 \, {\rm kg \, day}$ (i.e. a 100 kg
detector operating for a period of 2-3 years), assuming a local WIMP
density of $\rho_0\sim0.3 \, {\rm GeV \, cm}^{-3}$. For a detector only
capable of measuring the recoil axes the sensitivity would be reduced
by roughly an order of magnitude. We caution once more that these
numbers are for a detector with perfect recoil resolution and hence
provide an upper limit on the sensitivity of a real detector. If the
isotropy tests were only applied to the raw angle distributions these
numbers would be significantly larger, $~\sim 50-100\%$ for vector
data and an order of magnitude for axial data.

\subsection{Test for rotational symmetry}
\label{rotsymtest}

We now examine the number of events which would be required to detect
the deviation from rotational symmetry, $N_{\rm rot}$, for models B
(triaxial halo) and C (standard halo) using the Wilcoxon signed-rank
statistic (Appendix \ref{rotsymtest:A}), which is applied to the
reduced angle distributions. For both models and all detector
locations the reduced angle distributions in the meridian plane are
completely symmetric.  At Boulby the deviations from symmetry for
model B in the other two planes are also small and $>20\, 000$ events
would be required. This is also the case for the east-west plane at
Kamioka, but for the other location/read-out plane configurations the
deviation from symmetry could be detected with of order $5\, 000$
events.  The difficulty of detecting flattening of the halo is not
surprising as even with 3-d read-out and extremely flattened halo
models thousands of events were required~\cite{pap1}.
For model C the number of events is
strongly dependent on the read-out plane and detector location and
reflects the size of the mean reduced angle in each plane.  The
direction of the stream is close to perpendicular to the direction of
Solar motion, so it is not surprising that the plane which is best for
detecting isotropy (meridian) is the worst for detecting the stream.
For the $10^5 \, {\rm kg \, day}$
exposure considered in the previous subsection, rotational symmetry
could be rejected at $90\%$ confidence down to $\sigma_0\sim 1, 10$
and $>50 \times10^{-7} \, {\rm pb}$ for $\rho_0\sim0.3 \, {\rm GeV \,
cm}^{-3}$ for model C with a detector located at Boulby.
This test is not applicable to axial data.

\begin{table}[htbp]
\begin{tabular}{|c|c|c|c||c|c|c|}
\hline
Halo & \multicolumn{3}{c||}{ 
 $( R_{\rm c}, A_{\rm c})=(0.90,0.90)$} &
\multicolumn{3}{c|}{ 
 $( R_{\rm c}, A_{\rm c})=(0.95,0.95)$}
\\
\cline{2-7}
Model &  Boulby & Sudbury & Kamioka 
& Boulby & Sudbury & Kamioka
\\

\hline \hline
 & \multicolumn{6}{|c|} {Horizon plane} \\
\hline
B & $>$20 000 & 4 800 & 5 700 & $>$20 000 & 7 700 & 8 900  \\
C & 490  &  1 500 & 4 500 & 760 & 2 300  &  6 900 \\
\hline
 & \multicolumn{6}{c|} {Meridian plane} \\
\hline
B &  $>$20 000 & $>$20 000 & $>$20 000 & $>$20 000 & $>$20 000 & $>$20 000   \\
C & $>$20 000  & $>$20 000 & $>$20 000 & $>$20 000  & $>$20 000 &  $>$20 000  \\
\hline
 & \multicolumn{6}{c|} {East-west plane} \\
\hline
B &  $>$20 000 & 5 800& $>$20 000 &   $>$20 000  & 9 100& $>$20 000   \\
C & 4 400 & 1 800 &  461 & 6 900 & 2 800& 740\\
\hline 
\end{tabular}
\caption{Number of recoil events
required to reject rotational symmetry of recoil directions, 
$N_{\rm rot}$, at $90  \, (95)\%$
confidence in $90 \, (95)\%$ of experiments 
for the Wilcoxon signed rank statistic (appendix A3) for
halo models B (triaxial) and C (standard halo plus stream) 
and each read-out plane and detector location.}
\label{neventsrot}
\end{table}

\subsection{Test for mean direction}
\label{dirtest}

Finally we use the Watson mean direction test
(Appendix~\ref{meandirtest:A}) to find the number of events required
to detect the deviation from zero of the mean reduced angles for halo
model C. The numbers of events required are, for a detector located at
Boulby, $\sim 2.5, 10$ and $>10^{2}$ times those required with 3-d
readout~\cite{pap1} and are similar for the horizon plane and
significantly ($\sim 50\%$) smaller in the east-west plane  than
those required by the rotational symmetry test. This shows that
the mean direction test is more powerful than the rotational symmetry
test for detecting a tidal stream.  For a $10^5 \, {\rm kg \, day}$
exposure considered in the previous subsection, zero mean reduced
angle could be rejected at $90\%$ confidence down to $\sigma_0\sim 1,
6$ and $>50 \times10^{-7} \, {\rm pb}$ for $\rho_0\sim0.3 \, {\rm GeV
\, cm}^{-3}$ with a detector located at Boulby.
This test can not be applied to axial data.

\begin{table}[htbp]
\begin{tabular}{|c|c|c|c||c|c|c|}
\hline
Halo & \multicolumn{3}{c||}{ 
 $( R_{\rm c}, A_{\rm c})=(0.90,0.90)$} &
\multicolumn{3}{c|}{ 
 $( R_{\rm c}, A_{\rm c})=(0.95,0.95)$}
\\
\cline{2-7}
Model &  Boulby & Sudbury & Kamioka 
& Boulby & Sudbury & Kamioka
\\

\hline \hline
 & \multicolumn{6}{|c|} {Horizon plane} \\
\hline
C & 470  & 980  & 2 500 & 710 & 1 500 & 3 800 \\
\hline
 & \multicolumn{6}{c|} {Meridian plane} \\
\hline
C & $>$20 000 & $>$20 000 & $>$20 000 & $>$20 000   & $>$20 000 & $>$20 000 \\
\hline 
 & \multicolumn{6}{c|} {East-west plane} \\
\hline
C & 2 700   & 1 300& 490 & 4 200  & 1 900 & 750 \\
\hline
\end{tabular}
\caption{Number of recoil events
required to detect a deviation of the mean direction
from the direction of solar motion, 
$N_{\rm dir}$, at $90  \, (95)\%$
confidence in $90 \, (95)\%$ of experiments 
using the Watson mean direction test (appendix~\ref{meandirtest:A}) for
halo model C (standard halo plus tidal stream) for each read-out plane
and detector location.}
\label{neventsmean}
\end{table}

We caution that model C has parameter values at the optimistic ends of
the ranges estimated in Ref.~\cite{gondolo:sgr2}, i.e. high density
and low velocity dispersion. A lower stream density and/or a higher
velocity dispersion would give a peak recoil direction closer to the
mean direction of motion of the Sun, and make the deviation due to the
stream harder to detect. In general the directional detectability of
cold streams of WIMPs will clearly depend on how much the projection
of their bulk velocity into the read-out plane deviates from the
projection of the direction of solar motion

\section{Discussion}
\label{discuss}

We have studied the application of non-parametric tests, developed for
the analysis of circular data~\cite{mardia:jupp,fisher}, to the
analysis of simulated data as expected from a TPC-based directional
WIMP detector with 2-d read-out. As the (energy and direction
dependent) resolution of a 2-d directional detector has not been
calculated to date we assume perfect resolution. Our results therefore
provide a lower limit on the number of events required with a real
detector.

We found that if the senses of the recoils are known then between 10
and 30 events, depending on the read-out plane, will be required to
reject isotropy of the reduced (with the direction of motion of the
Sun subtracted) angle distribution. If the senses are not known then
these numbers are increased by roughly an order of magnitude. These
numbers are broadly similar to those for full 3-d read-out. If the
isotropy tests are applied to the raw angle distribution, however,
these numbers increase significantly; $\sim 50 \%$ in the meridian and
east-west planes and a factor of 3 in the horizon plane for vectorial
data. For axial data the increase is even larger, at least an order of
magnitude. Using the reduced angle distribution also has the advantage
that the time-dependence of the transformation means that even anisotropic
lab backgrounds will have isotropic reduced angle distributions.
It is therefore crucial that recoil events in a detector
with 2-d read-out are accurately time stamped and the reduced angles
calculated and analysed. The number of events required is always smallest
in the meridian plane, which contains the Earth's spin axis, as for this 
plane the 2-d projection effects which reduce the size of the isotropy 
are minimised.

After rejecting isotropy the next step would be to study the direction
dependence of the signal and attempt to derive information about the
dark matter distribution.  If a significant fraction of the local dark
matter distribution is in the form of a cold flow/tidal stream then
the peak recoil distribution will deviate from the direction of solar
motion, or in other words the mean reduced (with the direction of
solar motion subtracted) angle will differ from zero. The size of the
mean reduced angle (and hence the detectability of the stream) depends
on the direction and density of the stream. As an example we consider
a stream with bulk velocity, in Galactic co-ordinates, $(-65.0, \,
135.0 \, -250.0) \, {\rm km \, s^{-1}}$ comprising $25\%$ of the local
density~\cite{helmi,gondolo:sgr2}. 
The number of events required depends on the size of the mean reduced angle
and hence the detector location and read out plane. In the meridian
plane (which was best for rejecting isotropy)
the mean reduced angle is essentially zero, and the stream can not be 
detected, for all detector locations. For the other read-out planes
the number of events ranges between  500 and $5\,000$.
In the horizon (east-west) plane
the deviation of the mean reduced angle from zero decreases (increases)
as the detector location is moved South from Boulby to Kamioka and the
number of events required hence increases (decreases).
The number of events required could in theory be reduced by
optimising the choice of read-out plane, however in reality this is
not feasible due to technical limitations and our lack of knowledge of
the underlying WIMP distribution.

It is also potentially interesting to look for deviations from
rotational symmetry due to either flattening of the Milky Way halo or
the presence of a tidal stream. Unfortunately only a very significant
flattening of the Milky Way halo would be detectable. A tidal stream
could be detected in this way, however the number of events required
is larger than for the mean direction test.

In a realistic directional detector it will not be possible to measure
the nuclear recoil direction with infinite precision due to multiple
scattering and diffusion.  As discussed in Sec.~\ref{recoil} the
angular resolution of a 2-d detector, which will be a function of
recoil energy and primary direction due to projection effects, has not
yet been calculated.  We have therefore assumed a detector with
perfect resolution throughout.  A rough estimate of the likely
degradation in performance due to finite resolution can be obtained by
examining the fraction of recoils retaining a sufficiently large
($>80\%$) fraction of their length after 2-d projection so that their
direction can be reconstructed. For the standard halo model $57\%$,
$63\%$ and $60\%$ of the recoils, in the horizon, meridian and
east-west planes respectively, retain $>80\%$ of their length.
Therefore the numbers of events required for a realistic 2-d detector
with finite direction resolution are likely to be at least a factor of
order 2 larger than the numbers we obtain.

In summary, we have found that if the sense of the recoils can be
measured, then the potential for detecting a WIMP signal (via its
anisotropy) with a detector with 2-d read-out is similar to that for a
detector with full 3-d read-out, provided the reduced angle
distribution is utilised. If the senses of the recoils can not be
measured then the number of events required to detect the anisotropy
of a WIMP signal is increased by an order of magnitude, which again is
similar to the case of 3-d read out. We should caution, however, that
this comparison assumes a detector with perfect resolution. The
degradation in performance due to the finite resolution of a real
detector might be more significant for 2-d read out than for 3-d read
out.

\begin{acknowledgments}

A.M.G and B.M are supported by PPARC.

\end{acknowledgments}

\appendix

\section{Statistical tests for circular data}
\label{stat:A}

\subsection{Introduction}
\label{intro:A}

We first introduce the necessary basic definitions and terminology
relating to circular statistics. For further
information see the standard textbooks on this area~\cite{mardia:jupp,fisher}.

2-d vectors are most easily parameterised via their angles $\phi$ (relative
to some arbitrary fiducial direction).
Given a sample of $n$ 2-d vectors $\phi_{1},.... \phi_{n}$,
if we define 
\begin{eqnarray} 
C &=& \sum_{i=1}^{n} \cos{\phi_{i}} \,, \\ 
S &=& \sum_{i=1}^{n} \sin{\phi_{i}} \,, 
\end{eqnarray} 
then the resultant length of the sum of the vectors, $R$, is
given by $R^{2}= C^2 + S^2$ and the mean direction $\bar{\phi}$ can
be calculated via 
\begin{equation} 
\bar{\phi} = \arctan{(S/C)} \,, 
\end{equation}
adding $\pi$ if $C<0$ and $ 2 \pi$ if $S<0, C>0$.

The Cartesian coordinates of
the centre of mass are denoted by $(\bar{C}, \bar{S})$ 
where $\bar{C}= C/n$ and 
$\bar{S}= S/n$, and the mean resultant length is given by 
$\bar{R}= ( \bar{C}^2 + \bar{S}^2 ) ^{1/2}$.

With axial data (i.e. unsigned lines) the standard
procedure~\cite{axes,fisher} is to double the axial angles, reduce
them modulo $360^{\circ}$ and analyse the resulting vector data. There
is no rigorous justification for his procedure and it is somewhat
limited in its scope~\cite{fisher}.

\subsection{Tests of uniformity}
\label{iso:A}

The simplest test for uniformity is the Rayleigh test, which uses the mean 
resultant length, $\bar{R}$, which should be zero, modulo statistical
fluctuations, for angles drawn from a uniform distribution. The
modified
Rayleigh~\cite{rayleigh} statistic ${\cal R}^{\star}$, 
defined as~\cite{cf,mardia:jupp}
\begin{equation}
{\cal R}^{\star} = \left( 1 - \frac{1}{2n} \right) 2 n \bar{R}^2 +
        \frac{n \bar{R}^4}{2} \,.
\end{equation}
has the advantage of approaching
its large $n$ asymptotic distribution for smaller values of $n$ than $\bar{R}$.
Under the null hypothesis that the distribution
from which the sample of angles is drawn is isotropic, ${\cal R}^{\star}$
is asymptotically distributed as $\chi_{2}^{2}$ 
with error of order $n^{-1}$~\cite{mardia:jupp}.
This test is generally powerful, but is not sensitive to anisotropic 
distributions with zero mean resultant length (such as antipodally symmetric
distributions).

\vspace{0.2cm}

The Kuiper test~\cite{stephens,mardia:jupp,kuiper} is a variation of the
well-known Kolmogorov-Smirnov test which measures the maximum deviation
between the sample cumulative distribution function (cdf) and the cdf 
of the uniform distribution.
The Kuiper test has the advantage of being
invariant under cyclic transformations and equally sensitive to
deviations between the cdfs over the entire range of $\phi$.
As in the case
of spherical (i.e. 3-d) data the modified Kuiper statistic is defined as 
\begin{equation}
{\cal V}^{\star} = {\cal V} \left(n^{1/2} +0.155+ 
     \frac{0.24}{n^{1/2}}\right) \,,
\end{equation}
where ${\cal V}$ is the (unmodified) Kuiper statistic~\cite{kuiper,mardia:jupp} 
\begin{equation}
{\cal V} = {\cal D}^{+} + {\cal D}^{-} \,,
\end{equation}
and
\begin{eqnarray}
{\cal D}^{+}  &=& {\rm max} \left( \frac{i}{n} - U_{i} \right),\;\;\;
i=1,\ldots,n\\
{\cal D}^{-} &=& {\rm max}\left( U_{i} - \frac{i-1}{n} \right) \,,
\end{eqnarray}
where $U_{i} = \phi_{i} / 2
\pi$ and the $U_{i}$ have been ordered so that $U_{j}  \leq U_{j+1}$.
An analytic expression for the asymptotic distribution of 
${\cal V}^{\star}$ under the null hypothesis of uniformity is not available, 
so we calculate the null distribution numerically via Monte Carlo simulation.

\vspace{0.2cm}

Another alternative test uses Watson's ${\cal U}^2$ statistic 
which measures the mean square deviation between the sample cdf and the 
cdf of the uniform distribution.
The modified $U^{2}$ statistic~\cite{stephens,mardia:jupp} is defined as
\begin{equation}
{\cal U}^{\star 2} = \left( {\cal U}^{2} - 
      \frac{0.1}{n} +\frac{0.1}{n^2} \right)
       \left(1 + \frac{0.8}{n} \right) \,,
\end{equation}
 where ${\cal U}^2$ is Watson's statistic~\cite{watson,mardia:jupp} 
which can be written as
\begin{eqnarray}
{\cal U}^{2} &=& \sum_{i=1}^{n} \left( U_{i} - \bar{U} -
       \frac{i- 1/2}{n} + \frac{1}{2} \right)^{2} + \frac{1}{12 n} 
        \,,  \nonumber \\
         &=&
       \sum_{i=1}^{n} U_{i}^2  - n\bar{U}^2 
        - \frac{2}{n} \sum_{i=1}^{n} i U_{i} \nonumber \\
  &+& 
 (n+1) \bar{U} + \frac{n}{12} \,,
\end{eqnarray}
with $\bar{U} = \sum_{i=1}^{n} U_{i}/n$. 
As with the Kuiper statistic, the null distribution of ${\cal U}^{\star 2}$
has to be calculated numerically.

\subsection{Tests for rotational symmetry}
\label{rotsymtest:A}

The Wilcoxon signed-rank statistic~\cite{wilcoxon} can be used to test
for symmetry about a given direction $\mu_{0}$.  The data is first
rotated, i.e. $\phi_{i} \rightarrow \phi_{i} = \phi_{i} - \mu_{0}$,
and the $\phi_{i}$ ordered so that $\phi_{j} < \phi_{j+1}$. The rank
of each $|\phi_{i}|$ amongst $|\phi_{1}|$, ..., $|\phi_{n}|$ is
calculated and the test statistic ${\cal W}_{n}^{+}$ is given by the sum
of the ranks corresponding to $\phi_{i} > 0$. 
Any $\phi_{i}=0$ should
be removed from the data and the sample size $n$ reduced
correspondingly, and if more than one $\phi_{i}$ has the same absolute
value, then they should each be assigned the corresponding average
rank.

For $n>16$
\begin{equation}
{\cal W}^{+ \star} = \frac{{\cal W}_{n}^{+} - n(n+1)/4 +0.5}{\left[
     n(n+1)(2n+1)/24 \right]^{1/2}} 
\end{equation}
is normally distributed~\cite{lehmann}.

\subsection{Tests for mean direction}
\label{meandirtest:A}

A simple test for a given mean direction
can be preformed using Watson's ${\cal S}$ statistic which directly measures
the deviation of the mean of the sampled angles ($\bar{\phi}$) 
from the hypothesised mean direction ($\mu_{0}$)
~\cite{watson2,fisher}. The statistic is defined as
\begin{equation}
{\cal S} = \frac{\sin ( \bar{\phi} - \mu_{0}) }{\hat{\sigma}} \,,
\end{equation}
where
\begin{equation}
\hat{\sigma} = \frac{1}{ \sqrt{2} n \bar{R} } \left[
        n - \sum_{i=1}^{n} \cos{ 2 (\phi_{i} 
   - \bar{\phi})} \right]^{1/2} \,,
\end{equation}
is the sample circular standard error.
For $n \geq 25$ ${\cal S}$ is normally distributed under the 
null hypothesis that the sample is drawn from a distribution with mean 
direction $\mu_{0}$.


\end{document}